# B(E2) value of even-even $^{108-112}$Pd isotopes by interacting boson model-1[*]

I. Hossain[1†], H. Y. Abdullah[2], I. M. Ahmad[3], M. A. Saeed[4]

[1]Department of Physics, Rabigh College of Science and Arts, King Abdulaziz University, Rabigh 21911, Saudi Arabia

[2]Department of Physics, College of Education, Scientific Department, Salahaddin University, Erbil, Krg, Iraq

[3]Department of Physics, College of Education, Mosul University, 966 Mosul, Iraq

[4]Department of Physics, Universiti Teknologi Malaysia, 81310 Skudai, Johor, Malaysia

**Abstract-** This work studies the systematic reduced transition probabilities B(E2)↓, intrinsic quadrupole moments and deformation parameters of Pd isotopes with even neutrons from N= 62 to 66. The downward reduced transition probabilities B(E2)↓ from gamma transition $8^+$ to $6^+$, $6^+$ to $4^+$, $4^+$ to $2^+$ and $2^+$ to $0^+$ states of even-even $^{108-112}$Pd isotopes were calculated by the Interacting Boson Model (IBM-1) and compared with the available previous experimental results. The ratio of the excitation energies of the first $4^+$ and the first $2^+$ excited states, $R_{4/2}$, is also studied for the classification of symmetry of these nuclei. Furthermore we have studied systematically the transition rate R = B(E2: $L^+ \to (L-2)^+$ )/ B(E2: $2^+ \to 0^+$) of some of the low-lying quadrupole collective states in comparison with the available experimental data. The associated quadrupole moments and deformation parameters have been calculated. The results of this calculation are in good agreement with the corresponding available experimental data. The $^{108-112}$Pd isotopes show the O(6) symmetry.

**PACS-** 23.20.-g, 42.40.Ht, 42.30.Kq

## 1 Introduction

The Interacting Boson Model-1 (IBM-1) of Arima and Lanchello [1,2] has been widely accepted in describing the collective nuclear structure by the prediction of the low-lying states and the description of electromagnetic transition rates in the medium mass nuclei. In the first approximation, only pairs with angular momentum L =0 (called s-bosons) and L= 2 (called d-bosons) are considered. In the original form of the model known as IBM-1, the proton and neutron boson degrees of freedom are not distinguished. The model is associated with an inherent group structure, which allows for the introduction of limiting symmetries called *U(5)*, *SU(3)* and *O(6)* [1-3].

The yrast states up to $I^\pi= 8^+$ in Z= 46 isotones were found by $\pi g_{9/2}^{-4}$ configurations for the Z=50 closed shell. The low-lying collective quadrupole E2 excitations occur in even-even nuclei Z=46, which have been studied both theoretically and experimentally [4-8]. These investigations have intensively explored the nuclear shape degree of freedom of neutron-rich Pd isotopes by analyzing the rotational band structures, and a possible shape transition from a triaxial probate to a triaxial oblate [9,10]. The absolute transition probabilities for the first $2^+$ states in $^{110,114}$Pd have also been measured using the recoil distance Doppler shift technique following projectile Coulomb excitation at intermediate beam energies [11] and a sizable deformation in $^{114}$Pd has been determined, which is consistent with systematic IBM-1 calculations [12,13].

---

[*] Authors would like to thanks King Abdulaziz University for supporting this work. The authors would like to thank Ministry of Higher Education (MOHE) Malaysia/Universiti Teknologi Malaysia (UTM) for supporting this research work.

[†] Corresponding author, E-mail: hossain196977@yahoo.com

There is a number of theoretical work discussing intruder configuration and configuration mixing within framework of IBM-1 around the shell closure Z=50. For instance, the empirical spectroscopic study within the configuration mixing calculation in IBM [14,15], the IBM configuration mixing model in strong connection with the shell model [16,17], the conventional collective Hamiltonian approach [18,19] and the one starting from self-consistent mean-field calculation with microscopic energy density functional [20]. The collective properties of the $^{106-110}$Pd isotopes with quadrupole band structure are distributed by the influence of intruder states [4]. Recently we have studied the evolution properties of the yrast states for even-even $^{100-110}$Pd isotopes [21]. The electromagnetic reduced transition probabilities of even-even $^{104-112}$Cd isotopes were studied by Abdullah et al. [22].

In even-even nuclei, the reduced E2 probability B (E2; $0^+_1 \rightarrow 2_1^+$) from $0^+$ ground state to the first-excited $2^+$ state [23] is especially important, and for a deformed nucleus this probability (denoted here by $B(E2\uparrow)$) depends on the magnitude of the intrinsic quadrupole moment (quadrupole moment of the intrinsic state of the nucleus) and, hence, on deformation. At present the main purpose of this study is to evaluate the *B(E2)* value of ground state band through *E2* transition strengths, deformation parameter, intrinsic and various levels of quadrupole moment of even $^{108-112}$Pd isotopes within the framework of *IBM-1*.

## 2. Theoretical calculation

### 2.1. Reduced transition probabilities *B(E2)*

To calculate the *B(E2)* value the reduced matrix elements of the E2 transition operator ($T^{E2}$) have the form [1]

$$T^{E2} = \alpha_2 [d^*s + s^*d]^{(2)} + \beta_2 [d^*d]^{(2)} \qquad (1)$$

where $\alpha_2$ is the role of effective boson charge and $\beta_2$ is a parameter related to $\alpha_2$. The low-lying levels of even-even nuclei ($J_i$ =2, 4, 6, 8…….) usually decay by *E2* transition to the lower-lying yrast level with $J_f = J_i - 2$. The reduced transition probabilities in *IBM-1* are given for the limit *U(5)-O(6)* [24].

$$B(E2; L+2 \rightarrow L)\downarrow = \frac{1}{4}\alpha_2^{\,2}(L+2)(2N-L) \qquad (2)$$

where *L* is the state that nucleus translates to and *N* is the boson number, which is equal to half the number of valence nucleons (proton and neutrons). From the given experimental value of transition ($2^+ \rightarrow 0^+$), one can calculate the value of the parameter $\alpha_2^2$ for each isotope, where $\alpha_2^2$ indicates square of the effective charge. This value is used to calculate the transition $8^+$ to $6^+$, $6^+$ to $4^+$, $4^+$ to $2^+$ and $2^+$ to $0^+$.

## 2.2. Quadrupole moments

The intrinsic quadrupole moments ($Q_0$) of nuclei can be derived [25].

$$Q_0 = \frac{3}{\sqrt{5\pi}} Z R_0^2 \beta \qquad (3)$$

The quadrupole moment $Q(J)$ is related to $Q_0$ [25].

$$Q(J) = \frac{3k^2 - J(J+1)}{(J+1)(2J+3)} Q_0 \qquad (4)$$

and in the considered ground state $J=k$,

$$Q(J) = \frac{J(2J-1)}{(J+1)(2J+3)} Q_0 \qquad (5)$$

## 2.3. Deformation parameters

The upward electromagnetic quadrupole transition probability $B(E2)\uparrow$ is related to this value [26],

$$B(E2; J_i \rightarrow J_f)\downarrow = B(E2, J_f \rightarrow J_i)\uparrow \times g \qquad (6)$$

and

$$g = \frac{(2J_f + 1)}{(2J_i + 1)} \qquad (7)$$

The value of $B(E2)$ in units of $e^2b^2$, is related to $B(E2)$ in units of Weisskopf single particle transition (W.u) [27].

$$B(E2)\ e^2b^2 = 5.94 \times 10^{-6} \times A^{4/3} \times B(E2)\ w.u. \qquad (8)$$

Here e is the charge of electron and b (1 barn = $10^{-28}$ square meters) is the unit of area.

In vibration model the probability $B(E2)\uparrow$ of transition $0_1^+$ to $2_1^+$ is connected with the squared deviation $\beta^2$ [28] of nucleus shape from equilibrium as

$$B(E2; 0^+ \rightarrow 2^+)\uparrow = [(3/4\pi)eZ R_0^2]^2 \beta^2 \qquad (9)$$

The quadrupole deformation parameter $\beta$ can be calculated [28]

$$\beta = [B(E2)\uparrow]^{1/2} [3ZeR_0^2/4\pi]^{-1} \qquad (10)$$

where Z is the atomic number, and $R_0$ is the average radius of nucleus

$$R_0^2 = 0.0144\ A^{2/3}\ b \qquad (11)$$

A is the mass number of a nucleus.

## 3. Results and discussion

The boson number, transition level and downward electric quadrupole reduced transition probabilities $B(E2)\downarrow$ for the ground state band from $8^+$ to $6^+$, $6^+$ to $4^+$, $4^+$ to $2^+$, and $2^+$ to $0^+$ of even-even $^{108-112}$Pd isotopes are presented in Table 1. A boson represents the pair of valence nucleons and boson number is counted as the number of collective pairs of valence nucleons. At present $^{132}$Sn doubly-magic nucleus is taken as an inert core to find the boson number and it is presented in Table 1. Using known experimental $B(E2)\downarrow$ from $2_1^+ \to 0_1^+$ transition, the reduced transition probabilities of $4_1^+ \to 2_1^+$, $6_1^+ \to 4_1^+$ and $8_1^+ \to 6_1^+$ transitions of even-even $^{108,110,112}$Pd isotopes are calculated using *IBM-1* and presented in Table 1. The calculated results are also compared with the previous experimental results [28-31].

Table 1. Reduced transition probability $B(E2)\downarrow$ in even $^{108-112}$Pd nuclei

| Nuclei | Boson # | $I^+$ | $E_{exp}(I)$ KeV | Transition level | $E_\gamma$ KeV | $B(E2)_{Ref}$ W.U. | $B(E2)_{IBM-1}$ W.U. |
|---|---|---|---|---|---|---|---|
| $^{108}$Pd | 8 | 2 | 433.94 | $2^+ \to 0^+$ | 433.94 | 49.5(13) | 49.5(13) |
|  |  | 4 | 1048.25 | $4^+ \to 2^+$ | 614.28 | 73(8) | 88.63(228) |
|  |  | 6 | 1771.16 | $6^+ \to 4^+$ | 722.91 | 107(11) | 111.38(293) |
|  |  | 8 | 2548.42 | $8^+ \to 6^+$ | 777.20 | 148(20) | 123.76(326) |
| $^{110}$Pd | 9 | 2 | 373.80 | $2^+ \to 0^+$ | 373.8 | 55.5(9) | 55.5(9) |
|  |  | 4 | 920.78 | $4^+ \to 2^+$ | 547.04 | 90(7) | 98.67(160) |
|  |  | 6 | 1574 | $6^+ \to 4^+$ | 653.10 | 108(11) | 129.51(210) |
|  |  | 8 | 2296 | $8^+ \to 6^+$ | 722.20 |  | 148.01(240) |
| $^{112}$Pd | 10 | 2 | 348.79 | $2^+ \to 0^+$ | 348.70 | 41(7) | 41(7) |
|  |  | 4 | 883.56 | $4^+ \to 2^+$ | 534.60 |  | 73.8(126) |
|  |  | 6 | 1551.3 | $6^+ \to 4^+$ | 667.90 |  | 98.4(168) |
|  |  | 8 |  | $8^+ \to 6^+$ |  |  | 114.8(196) |

Ref. [29-31]

Table 2 presents the calculation of the upward electric quadrupole reduced transition probabilities $B(E2)\uparrow$ of ground state band from $0_1^+ \to 2_1^+$, $2_1^+ \to 4_1^+$, $4_1^+ \to 6_1^+$, and $6_1^+ \to 8_1^+$ of even-even $^{108-112}$Pd isotopes. The intrinsic quadrupole moments $Q_0$, quadrupole moments $Q_J$ up to $8_1^+$ level, and deformation parameter $\beta$ correspond to the frame work of IBM-1 are presented and compared to the available experimental values. It is found that the calculated results for $B(E2)\uparrow, Q_0$ and $\beta$ values are consistent with the previous experimental results [28-31].

### 3.1 The R$_{4/2}$ classifications

The collective dynamics of the energies of even-even nuclei are grouped into classes, within each class the ratio of excitation energies of the first $4^+$ and the first $2^+$ excited states is:

$$R_{4/2} = \frac{E(4_1^+)}{E(2_1^+)}$$

The *R4/2* energy ratio is a fundamental observable to describe the nuclear structure. An harmonic vibrator U(5) has limit $E(4_1^+)/E(2_1^+)$ = 2.0 ~2.4, an axially symmetric rotor *SU(3)* should have $E(4_1^+)/E(2_1^+)$ =

3.0~3.3, O(6) should have $E(4_1^+)/E(2_1^+)$ = >2.4~2.7 and transitional nuclei have $E(4_1^+)/E(2_1^+)$ = >2.7 to < 3.0. The variation of the $E(4_1^+)/E(2_1^+)$ values as a function of even neutron numbers of Pd isotopes for experimental values [29-31], *U(5), O(6)* and *SU(3)* limits are presented in Fig.1. We identified *O(6)* symmetry in even-even nuclei with Z=46 and N = 62, 64, 66.

Table 2. Quadrupole moment and deformation parameter of even $^{108-112}$Pd nuclei

| Nucl. | Transition | B(E2)↑$_{Ref}$ e$^2$b$^2$ | B(E2)↑$_{IBM-1}$ e$^2$b$^2$ | β Ref. | β Cal. | $Q_{o\ Ref}$ (b) | $Q_o$cal (b) | *$Q_I$cal (b) |
|---|---|---|---|---|---|---|---|---|
| $^{108}$Pd | $0^+ \to 2^+$ | 0.761(23) | 0.755 | 0.243(6) | 0.242 | 2.76(7) | 2.75 | 0.79 |
| | $2^+ \to 4^+$ | 0.420(40) | 0.474 | | | | | 1.40 |
| | $4^+ \to 6^+$ | 0.470(50) | 0.489 | | | | | 1.73 |
| | $6^+ \to 8^+$ | 0.590(80) | 0.495 | | | | | 1.93 |
| $^{110}$Pd | $0^+ \to 2^+$ | 0.870(40) | 0.855 | 0.257(6) | 0.255 | 2.96(7) | 2.93 | 0.84 |
| | $2^+ \to 4^+$ | | 0.304 | | | | | 1.49 |
| | $4^+ \to 6^+$ | | 0.577 | | | | | 1.84 |
| | $6^+ \to 8^+$ | | 0.596 | | | | | 2.06 |
| $^{112}$Pd | $0^+ \to 2^+$ | 0.66(11) | 0.655 | 0.220(18) | 0.221 | 2.57(22) | 2.57 | 0.73 |
| | $2^+ \to 4^+$ | | 0.419 | | | | | 1.31 |
| | $4^+ \to 6^+$ | | 0.449 | | | | | 1.62 |
| | $6^+ \to 8^+$ | | 0.473 | | | | | 1.80 |

*$Q_I$=$2^+, 4^+, 6^+, 8^+$

Ref. [29-31]

### 3.2. Systematic reduced transition probabilities *B(E2)*

In order to obtain the value of reduced transition probabilities, we have fitted the calculated absolute strengths *B(E2)* of the transitions within the ground state band to the experimental ones. The value of the effective charge $(\alpha_2)$ of IBM-1 was determined by normalizing the experimental data $B(E2; 2_1^+ \to 0_1^+)$ of each isotope using Eq. (2). From the given experimental value of transitions ($2_1^+ \to 0_1^+$), we have calculated the value of the parameter $\alpha_2^2$ for each isotope and used this value to calculate the transitions from $4^+ \to 2^+$, $6^+ \to 4^+$ and $8^+ \to 6^+$. Fig. 2 shows the theoretical and experimental values of *B(E2)* in W.U. plotted as a function of transition levels. The results of the present work are compared with the previous experimental values [28-31] and are in good agreement within experimental error (Table 1). The present calculations are performed in the *U(5)-O(6)* limit and therefore a good agreement between the calculated values and the experimental ones indicated that Pd isotopes obey to this limit. The even-even $^{108-112}$Pd nuclei are nicely reproduced by the experimental data and their fits are satisfactory.

We have compared the ratio $R = B(E2: L^+ \to (L-2)^+ )/B(E2: 2^+ \to 0^+)$ of IBM-1 and the previous experimental values in the ground state bands (normalized to the $B(E2: 2^+ \to 0^+)$) as a function of angular momentum L and are shown in Fig.3. As a measure to quantify the evolution, it is shown that the results of *R* values increase with increasing the high spin states. We have found that the calculated values are in good agreement with the previous available experimental results [29-31]. Hence even-even 108-112Pd nuclei are O(6) symmetry. Actually, in IBM-1 the proton and neutron bosons are not distinguishable as long as valence protons and neutrons are both hole-like or both particle-like.

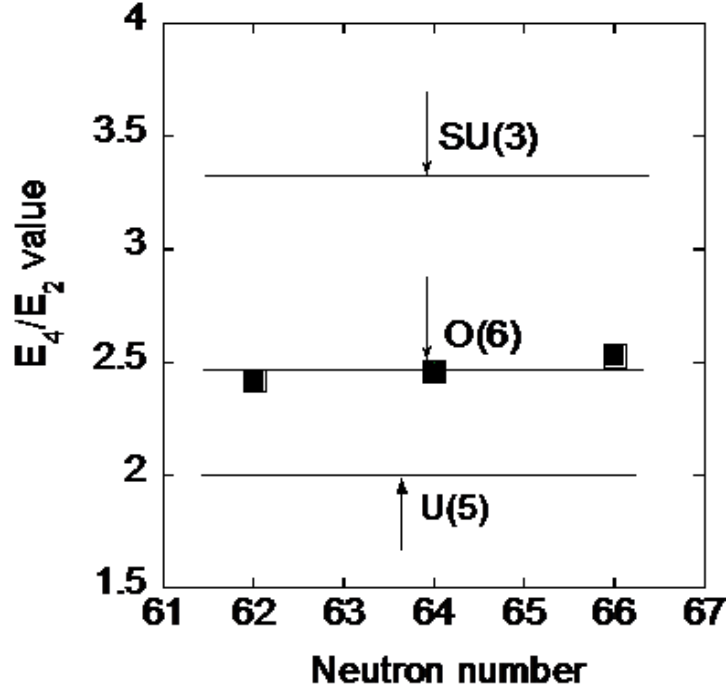

Fig.1. E $(4_1^+)$/ E$(2_1^+)$ in experimental values [29-31], U(5), O(6) and SU(3) limit of $^{108,110,112}$Pd isotopes.

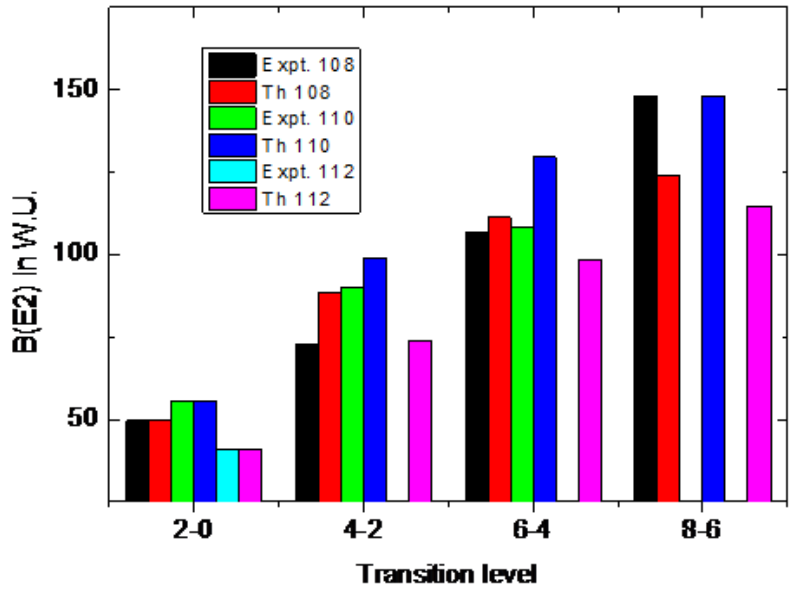

Fig. 2 Plot of B (E2; $I_i \rightarrow I_f$) values vs. ground state transition of angular momentum of $^{108,110,112}$Pd isotopes.

### 3.3. Quadrupole moments

The intrinsic quadrupole moments of $Q_0$ and $Q_J$ ($2^+, 4^+, 6^+, 8^+$) are calculated using Eqs. (3) and (5) respectively for even-even nuclei $^{108-112}$Pd and are presented in Table 2. The calculated intrinsic quadrupole moments corresponding to the frame work of IBM-1 were compared with the corresponding experimental values [28-31]. Fig. 4 shows the systematic theoretical values of $Q_0$ compared with the previous experimental results as a function of even neutron numbers for $^{108-112}$Pd isotopes. It is shown that the present values are in good agreement with the previous experimental results.

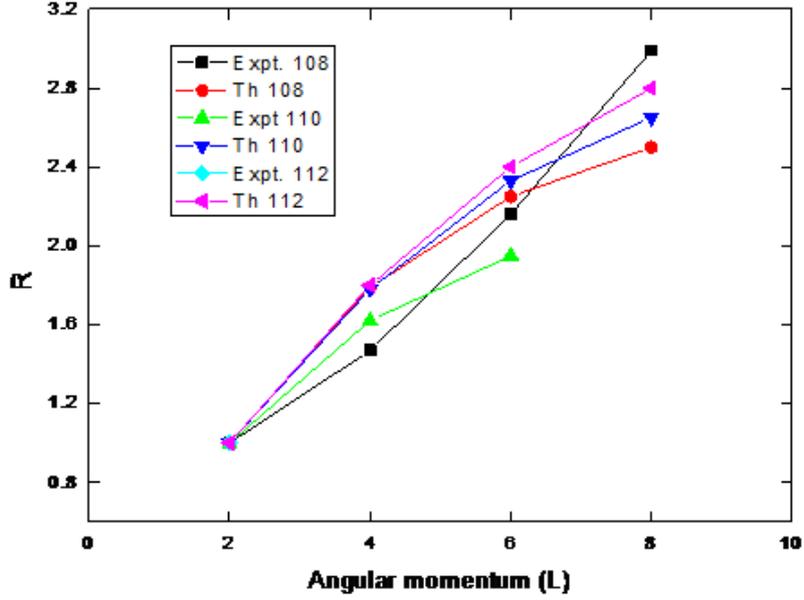

Fig.3. R values of $^{108,110,112}$Pd isotopes using IBM-1 and experiment [29-31]. The ratio R = B(E2: L$^+$→(L-2)$^+$)/B(E2: 2$^+$ → 0$^+$) in the ground state bands (normalized to the B(E2: 2$^+$ → 0$^+$)).

### 3.4. Deformation parameter ($\beta$)

The upward reduced transition probabilities $B(E2)\uparrow$ are obtained using Eqs. (6) and (7). The deformation parameters ($\beta$) of nuclei with proton Z=46 and even neutron N= 62 - 66 are obtained using Eq. (10) and presented in table 2. The calculated deformation parameters associated with the frame work of IBM-1 were compared with the corresponding previous experimental results [28]. Fig.5 shows the deformation parameter as a function of neutron number N. The present calculations are consistent with the corresponding calculations of previous experimental values.

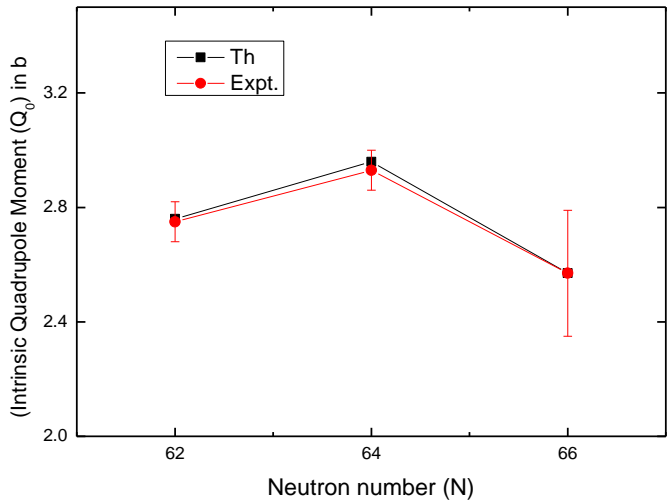

Fig. 4. The intrinsic quadrupole moment ($Q_0$) vs. the neutron number of $^{108,110,112}$Pd isotopes

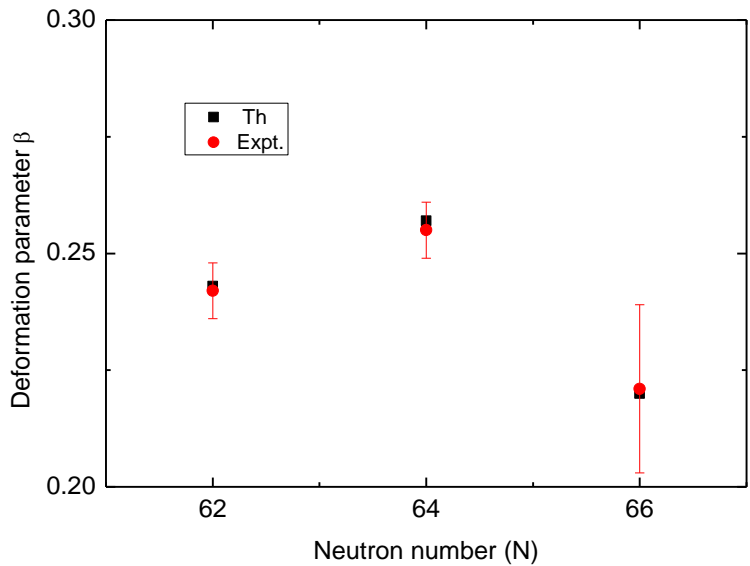

Fig. 5. The deformation β vs. the neutron number of $^{108,110,112}$Pd isotopes

## 4. Summary

We report the *B(E2)* values of even-even nuclei $^{108-112}$Pd by *IBM-1*. The associated quadrupole moments ($Q_0$ and $Q_J$) and deformation parameter ($β$) are also calculated. The calculated reduced transition probabilities, quadrupole moments and deformation parameters are consistent with the previous experimental results. The values of energy ratio R(4/2) of even-even $^{108-112}$Pd have been performed in the *O(6)* character. The results are very useful for compiling nuclear data table.